\documentclass[twocolumn,aps,ajp,groupedaddress]{revtex4}
\usepackage[pdftex]{graphicx}
\usepackage{graphics}
\usepackage{amsmath,amsthm,amsfonts,amssymb}
\usepackage{longtable}

\begin{document}
\title{Measured speed \emph{versus} true speed}
\author{Israel P\'erez }
\affiliation{Department of Applied Physics, CINVESTAV, M\'erida, Yucat\'an, M\'exico \\ Km 6 Ant. Carr. a Progreso Cordemex A.P. 73, C.P. 97305}
\email{cooguion@yahoo.com, iperez@mda.cinvestav.mx}

\begin{abstract}
The theoretical predictions, derived from the Lorentz and the Tangherlini transformations, for the one-way speed of any physical entity are confronted with the corresponding expressions for the one-way measured speed obtained from a gedanken experiment. The experiment demonstrates that, for an inertial frame $K'$ in motion relative to an inertial frame $K$ where the one-way speed of light is isotropic, even the special theory of relativity renders the one-way speed of light as function of the velocity of $K'$ in agreement with the Tangherlini transformations. However, the two-way speed of light remains constant for all inertial frames, in agreement with the two-way experimental techniques. This implies that there must exist \emph{one and only one} inertial frame where the one-way speed of light is isotropic. These investigations also show how we can determine, with certain restrictions, the true speed of a physical entity and of the true speed of $K'$ relative to $K$.
\end{abstract}
\maketitle

\section{Introduction}
\label{int}

From the historical point of view, it is well known that the special theory of relativity (SR) was split into two philosophies: that one supported by Einstein \cite{einstein2} and Minkowski \cite{minkowski}, which was based on the two postulates and considered as superfluous the existence of a material medium for the propagation of electromagnetic fields, and the other, upheld mainly by Larmor \cite{larmor1,larmor2}, Lorentz \cite{lorentz2} and Poincar\'e \cite{poincare3,poincare2}, which maintained the existence of a privileged frame. At the end, the lack of unambiguous experimental evidences of the privileged frame favored the former. 

A total turn, however, was given after the discovery of the cosmic background radiation and dark matter which implies that, cosmologically, a privileged frame of reference exists. As a consequence, an increasing number of physicists have questioned the philosophy \cite{mansouri,bell1,guerra1,guerra2,abreu,perez1} of SR. Besides of this, other weighty arguments against the validity of SR rest on the following experimental facts: (\emph{i}) The recent reanalysis of the aether drift experiments \cite{marinov1,munera,munera1,consoli,consoli1,consoli2,cahill,spavieri}. (\emph{ii}) The reflection of light beams at moving mirrors clearly suggests the idea of a privileged frame. The first postulate of SR violates Huyghens principle which is implied in this optical phenomenon \cite{perez1}. (\emph{iii}) It has not been experimentally possible to accurately synchronize two distant clocks. (\emph{iv}) In most (if not all) modern experimental techniques the electromagnetic fields must close circuits. This implies that most experiments are, actually, two-way experiments. The consequence of the last two points is that, for instance, it is, perhaps, impossible to test the one-way speed of light as predicted by the addition theorem of velocities in SR \cite{feenberg,guerra1,guerra2,abreu,spavieri,iyer1}.

In this contribution we shall endeavour to elucidate point (\emph{iv}). With the help of a particular gedanken experiment expressions for the one-way and two-way measured speed of any physical entity (PE) will be derived. 

\section{Preliminaries}
\label{preli}
\subsection{True and measured speeds}
One of the aims of the present investigation is to analyze the measuring processes and, consequently verify whether the experimental techniques allow us to know the ``\emph{true or  intrinsic}" speeds of the PEs. For convenience, we shall denote the ``\emph{measured speeds}" with a bar above the quantity, e.g. $\bar{v}$. The reason for this is just to make a clear epistemological distinction between these quantities. The features that distinguish the true speeds from the measured speeds will be logically assimilated as we advance. Furthermore, we shall restrict ourselves to study only \emph{direct} measurements of speed, i.e., by measuring space and time. Indirect measurements of speed like, for example, $c^2=\bar{E}/\bar{m}$, where $c$ is the speed of light in vacuum, $\bar{E}$ is the measured energy of a particle and $\bar{m}$ is the measured mass of a  particle, are outside of the scope of the present investigation. 

\subsection{A matter of semantics}
We should warn the reader that the subject may prove somewhat difficult to understand if we do not make clear the following peculiarity. It is well known, from the customary view, that the theoretical constant, say $V$, that appears in SR can be identified with the electromagnetic constant $c$ that appears in Maxwell's electrodynamics, i.e., the speed of light in vacuum. Indeed, the speed of light seems to be constant but only in the radiation zone \cite{jackson}.  The rigorous solutions of Maxwell's equations for the intermediate and near zones, however, do not tolerate that the speed of the electromagnetic fields be constant \cite{mugnai,budko}. Therefore, to avoid semantical misunderstandings \cite{ellis,mendelson}, we shall start our investigations assuming only the validity of the first postulate of SR. When one derives the Lorentz transformations from only the first postulate \cite{schroder,mermin,rindler} there remains a universal constant $V$, with the dimensions of velocity and of finite value, to be determined. To acquire physical meaning, the tacit condition that the transformation equations demand is that the speed $v$ of the moving inertial frame be less than $V$, otherwise the equations would become complex numbers. If the speed of light in the radiation zone is the limiting speed in the universe we can identify $V$ with $c$, so that the condition $v<V$ is not violated. Thus, for our purposes, $V$ will be assigned the value defined by Comit\'e International des Poids et Mesures (BIPM) \cite{giacomo0}, that is, $V \equiv 299\,792\,458$ m/s. Note that, although this value corresponds to $c$, it is a conventionally defined one. The conventional value could have been $299\,792\,459$ m/s or any other value. The relevance does not precisely reside on the magnitude but on the fact that it is assumed to be constant and maximal; the universality, in contrast, is founded on the principle of relativity. 

So far, we have assumed that the speed of light is constant and maximal in the radiation zone and in vacuum. This assumption is based on Maxwell equations and on the theoretical predictions of the principle of relativity. Nevertheless, we shall show below that a two-way experimental technique does not allow us to know neither the true speed of light nor the true speed of any PE unless certain conditions are satisfied. 

\section{The measured speed of any physical entity from a gendanken experiment}
\label{measmo}
To see the importance on how the experimental techniques influence the outcomes of a measurement, we shall deal with a two-way gedanken experiment to measure the speed of any PE. The experiment resembles one of the simplest and most typical two-way techniques used in a physics laboratory \cite{mugnai,budko,aoki,greaves}. And without loss of generality the same principles developed here can be applied for the analysis of any other two-way experimental technique.

\subsection{A non-trivial measurement}
First of all, it is assumed that the measurement is realized in an inertial frame $K$, where the one-way true speed of light $c$ is isotropic in the radiation zone and the Einstein's synchronization convention has been adopted. Following the jargon of Iyer and Prabhu \cite{iyer1}, this frame shall be called the \emph{isotropic frame}. 

\begin{figure}[htp]
\begin{center}
\includegraphics[width=3cm]{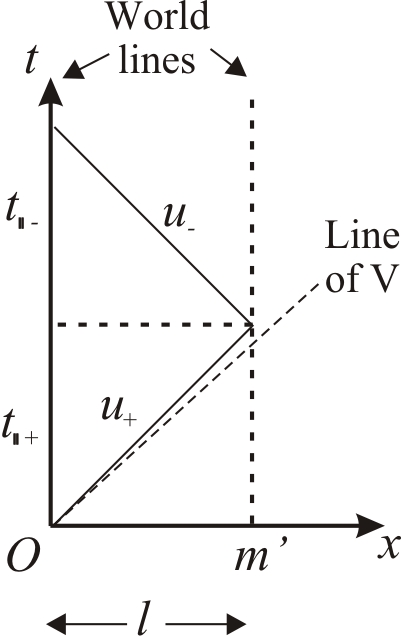}
\caption{Space-time diagram for the measurement of speed in the isotropic frame. The observer situated at the origin has to wait for the returning information that travels at true speed $u_-$ and thus he measures the total time $t_{\|}=t_{\|+}+t_{\|-}$.}
\label{isotro}
\end{center}
\end{figure}

Let us consider now that an observer in $K$, located at the origin $O$, is interested in measuring the true speed $u_{+}$ of a PE$_{\|+}$ by measuring the time it takes to travel an arbitrary distance $l$ (see Figure \ref{isotro}). 
How would the observer know that the PE$_{\|+}$ has arrived at the opposite endpoint? Certainly, the information of the arrival event has to return to the origin by any physical means at any true speed $u_{-}$. This can be achieved, for instance, by just observing the event (by means of a light signal), or putting a detector (by means of an electric field through a wire), etc. In any case, \emph{the observer has to wait for a returning information}. Imagine that at $t_{\|0}=0$ we let the PE$_{\|+}$ in question to depart from the origin $O$ and traverse the distance $l$ parallel to the spatial $x$ axis. At the opposite endpoint we place a device $m'$ that detects the PE$_{\|+}$ and returns the arrival information via any PE$_{\|-}$ (it could be the same PE) towards the spatial origin, where the observer, who is carrying a clock, measures the time $t_\|$ that it takes to complete the round trip. 

Bearing this in mind, the outward time or \emph{time of flight} of the PE$_{\|+}$ is $t_{\|+}=l/u_{+}$, and the time of the returning information or \emph{delay time} is $t_{\|-}=l/u_{-}$. By definition the speed of any PE is the space traversed by the entity divided by the total time spent in the motion. Thus the one-way measured speed for the PE$_{\|+}$ is 
\begin{equation}
\label{velk}
\bar{u}_{+}=l/(l/u_{+})=u_{+}.
\end{equation}
And for the returning information the one-way measured speed is
\begin{equation}
\label{velk}
\bar{u}_{-}=l/(l/u_{-})=u_{-}.
\end{equation}
It is clear that the measured speeds are equal to the true speeds. But the measurement is done only when the information returns to $O$, therefore, $t_{\|}=t_{\|+}+t_{\|-}$ and the two-way measured speed of the PE$_{\|+}$, actually, is 
\begin{equation}
\label{uaver}
\bar{u}=\frac{2l}{l/u_{+}+l/u_{-}}=\frac{2}{1/u_{+}+1/u_{-}}=\frac{2u_{+}u_{-}}{u_{+}+u_{-}}.
\end{equation}
Note that this expression clearly resembles the definition of the harmonic mean of the speed. It is to be seen, that the previous expressions are by no means trivial. Next we show why.
\begin{figure}[htp]
\begin{center}
\includegraphics[width=5cm]{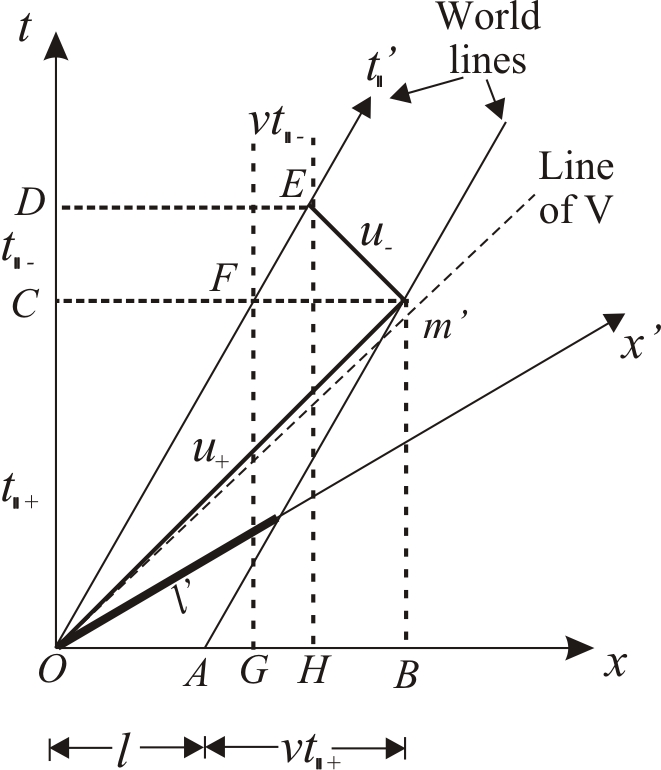}
\caption{Space-time diagram for the measurement of the speed realized in the frame $K'$ as seen by the isotropic frame.}
\label{stmeas}
\end{center}
\end{figure}

\subsection{Longitudinal motion}

Let us imagine that we wish to perform the measurement of the speed in an inertial frame $K'$ that is moving relative to $K$ in the $+x$ direction with true speed $v< V$. Figure \ref{stmeas} shows the space-time diagram for this problem. For simplicity we shall assume also that the measurement starts at $t_{\|}=t'_{\|}=0$. As judged from $K$, the PE$_{\|+}$ follows the path $Om'$, whereas the PE$_{\|-}$ follows the path $m'E$. From the figure we have that $u_+=OB/OC=OB/t_{\|+}$ and $u_-=HB/CD=HB/t_{\|-}$. Also $v=AB/OC=AB/t_{\|+}=GH/CD=GH/t_{\|-}$ and $OA=GB=l$. From these expressions we can derive the following relations
\begin{eqnarray}
\label{rmas2}  OB & = & u_+t_{\|+}=OA+AB=l+vt_{\|+}, \nonumber \\
\label{rmenos1} HB & = & u_-t_{\|-}=GB-GH=l-vt_{\|-}. \nonumber
\end{eqnarray}
Solving for the times from these relations we have that $t_{\|\pm}=l/(u_{\pm}\mp v)$. To determine the times spend in each journey, as judged in $K'$, we just have to consider length contraction $l=l'\gamma^{-1}$ and time dilation $t_{\|}=\gamma t'_{\|}$, where $\gamma=1/\sqrt{1-v^2/V^2}$. It follows that the observer in $K'$ measures
\begin{equation}
\label{tonpar}
t'_{\|+}=\gamma^{-2}\frac{l'}{u_{+}-v}; \qquad t'_{\|-}=\gamma^{-2}\frac{l'}{u_{-}+v}.
\end{equation}
Hence the one-way measured speeds are
\begin{equation}
\label{conpar}
\bar{u}'_{+}=\frac{l'}{t'_{\|+}}=\gamma^{2}(u_{+}-v); \qquad \bar{u}'_{-}=\frac{l'}{t'_{\|-}}=\gamma^{2}(u_{-}+v).
\end{equation}
Since the observer in $K'$ can only measure the round trip time then $t'_{\|}=t'_{\|+}+t'_{\|-}$ and the two-way measured speed of the PE$_{\|+}$ is
\begin{eqnarray}
\bar{u}'&=&\frac{2l'}{t'_{\|}}=  \frac{2l'}{\gamma^{-2}\frac{l'}{u_{+}-v}+\gamma^{-2}\frac{l'}{u_{-}+v}} \nonumber \\
\label{tottim} &=& \frac{2}{1/\bar{u}'_{+}+1/\bar{u}'_{-}}=\frac{2\gamma^{2}(u_{+}-v)(u_{-}+v)}{u_{+}+u_{-}}.
\end{eqnarray}
It is evident that the average round trip speed becomes the harmonic mean of the measured speed. Also note that this experimental procedure does not allow the observer in $K'$ to determine by himself the true speeds $u_{\pm}$ and $v$.

\subsection{Transversal motion}
\begin{figure}[htp]
\begin{center}
\includegraphics[width=5cm]{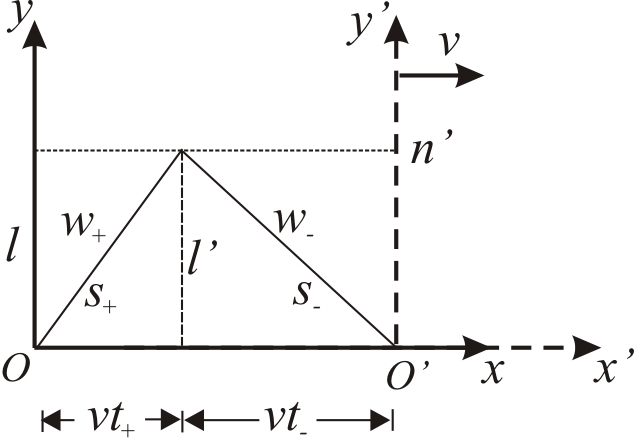}
\caption{Measurement conducted in $K'$ for the transversal motion as seen from the isotropic frame. Only the spatial dimensions are considered.}
\label{transmot}
\end{center}
\end{figure}

Now let us imagine that the motion of two other PEs$_{\bot \pm}$, in the outward and returning ways, takes place in $K$ along the $y$ axis. Such case is trivial and reduces to the longitudinal motion of the previous section. But when the experiment is conducted in $K'$ it acquires a distinct aspect. The spatial situation is depicted in Figure \ref{transmot}. 
As seen from $K$, the PE$_{\bot+}$, with true speed $w_+$, arrives at the point $n'$ from where the PE$_{\bot-}$, with true speed $w_-$, is guided towards the origin $O'$. To determine the time for each journey, we use the pythagorean theorem for the distances $S_{\pm}$ traveled by the entities during each journey. So, we have that $S_{\pm}^2=(w_{\pm}t_{\bot \pm})^2=(vt_{\bot \pm})^2+l^2$. On solving for the transversal times we obtain $t_{\bot \pm}=(l/w_{\pm})\gamma_{\pm}$, where $\gamma_{\pm}=1/\sqrt{1-v^2/w_{ \pm}^2}$. Note also that 
\begin{equation}
\label{wwe}
w_{\pm}=\sqrt{w_{x\pm}^2+w_{y\pm}^2}, \quad w_{x\pm}=v;  \quad w_{y\pm}=w_{\pm}\gamma^{-1}_{\pm}.
\end{equation}
Since the spatial dimension is perpendicular to the line of motion it follows that $l=l'$ and $t_{\bot}=t'_{\bot}\gamma$, hence
\begin{equation}
\label{otrt}
t'_{\bot \pm}=\frac{l'}{w_{ \pm}}\gamma_{\pm}\gamma^{-1}.
\end{equation}
Consequently, the one-way measured speeds in the transversal direction are
\begin{equation}
\label{uss}
\bar{w}'_{\pm}=w_{ \pm}\gamma^{-1}_{\pm}\gamma=\gamma w_{y\pm}.
\end{equation}
Whereas the two-way measured speed is given by
\begin{equation}
\label{twotrans}
\bar{w}'=\frac{2l'}{t'_{\bot}}=\frac{2}{1/\bar{w}'_++1/\bar{w}'_-}=\frac{2\gamma w_+w_-}{\gamma_+w_-+\gamma_-w_+}.
\end{equation}
Once again this is the harmonic mean of the speed in the transversal direction and the observer in $K'$ cannot solve for the variables $w_{\pm}$ and $v$. Let us see whether this trouble can be resolved.

\subsection{Finding the true speeds}
\label{truespe}
So far we have not imposed any restriction on the values of $u_{\pm}$ and $w_{\pm}$, nor on the temporal moments in which the longitudinal or transversal measurements are realized. We have only worked out the calculations for the longitudinal and transversal situations where we required that $v<V$. Under such scenario the observer in $K'$ still has five unknowns to find, namely: $u_{\pm}$, $w_{\pm}$, $v$. With the aid of Equations \eqref{tottim} and \eqref{twotrans}, however, two true speeds can be estimated if we constrain the experimental situation to the following conditions. (1) The medium for the propagation of the four PEs is isotropic and homogeneous and its temperature remains constant. (2) The previous point guarantees that, if we use the same PEs for the four paths, the speed of the PEs must be the same, hence $u\equiv w_{+}\approx w_{-}\approx u_{+}\approx u_{-}$. For instance, the physical entities could be electric fields traveling through ``identical wires". And (3) both measurements, longitudinal and transversal, are realized simultaneously. If these conditions are satisfied Equation \eqref{tottim} becomes
\begin{equation}
\label{asdf}
\bar{u}'=u\gamma^{2}\gamma^{-2}_u,
\end{equation}
whereas Equation \eqref{twotrans} reduces to  
\begin{equation}
\label{twotra}
\bar{w}'=u\gamma\gamma^{-1}_u.
\end{equation}
The previous equations give us hope to determine the true speed $u$ in terms of the measured quantities, that is
\begin{equation}
\label{sdfa}
u=\frac{\bar{w}'^2}{\bar{u}'}.
\end{equation}
The introduction of this expression into either of the equations \eqref{twotrans} or \eqref{asdf} also give us the speed $v$ of $K'$ relative to $K$ in terms of the measured quantities, namely
\begin{equation}
\label{adfe}
v=V\Biggl\{1-\frac{\bar{w}'^2}{\bar{u}'^2}\Biggl[1-\frac{\bar{w}'^4}{(\bar{u}'V)^2}\Biggr]\Biggr\}^{1/2}.
\end{equation}
It is to be noted, from this equation, that the value of $v$ depends on the definition of $V$. 

The arguments that we have given so far are in terms of speeds but they also hold if we consider only the times. In such case we just have to solve for the times from the equations of the previous sections. Thus, it is not difficult to foresee, that this gendaken experiment can be reproduced in the laboratory. The setup resembles a Michelson-Morley configuration, but in this case we just have to plug in the corresponding wires, forming a right angle, to the input channels of an oscilloscope and by measuring the times in each direction we can determine $u$ and $v$.

On the other hand, in the ideal event that $u=c=V$, that is, if we are sending forth and back light signals in vacuum, the two-way measured speed in any direction and $\bar{w}'_{\pm}$ become equal to $V$, however, $\bar{u}'_{\pm}$ remains velocity dependent in conflict with the prediction of SR (see next section). For this reason we shall call this frame the \emph{anisotropic frame}. If we follow this line of thought, then any frame in motion relative to the isotropic frame is also an anisotropic frame.

\section{Addition Theorem of Velocities}
\label{dis}

In 1977 Mansouri and Sexl \cite{mansouri}, and more recently, Iyer and Prabhu \cite{iyer1} derived a set of transformations that predicts the same length contraction and time dilation as the Lorentz transformations (LT). They showed that a transformation that retains absolute simultaneity is kinematically equivalent to SR. 
\begin{table}[htp]
  \centering 
  \caption{Relativistic Transformations}\label{transf}
  \begin{tabular}{cc}
\hline
   Lorentz (LT) &  Tangherlini (TT)\\ \hline
    $x'  =  \gamma(x-vt)$; &$x'  =  \gamma(x-vt)$;  \\
        $y'  =  y$; \quad $z'=z$; &$y'  =  y$; \quad $z'=z$;  \\
    $t'  =  \gamma(t-vx/V^2);$ &$t'  =  \gamma^{-1}t$.  \\
\end{tabular}
\end{table}
Table \ref{transf} shows the so-called Tangherlini transformations (TT) that only differ from the LT by the clock synchronization convention. Note that for $v\ll V$ both tranformations reduce to the Galilean  form.
From these transformations we can derive the addition theorem of velocities for each formalism, that is:
\begin{eqnarray}
\label {addnetgg}u'_{x} & = & \frac{u_{x}-v}{1-u_xv/V^2}; \quad u'_y= \frac{u_{y}\gamma^{-1}}{1-u_xv/V^2};  \quad \textrm{LT}\\
\label {addnetgg}u'_{x} & = & \gamma^2(u_{x}-v); \quad u'_y=\gamma u_y. \hspace{1.8cm}  \textrm{TT}
\end{eqnarray}
It is clear that the Equations \eqref{conpar} and \eqref{uss} agree with the latter of these expressions.

\section{Discussion}

In this contribution we have only dealt with one of the most representative two-way methods used to determine the speed of any PE. In our gedanken experiment the length $l_{+}=l$ for the outward motion is assumed to be equal to the length $l_{-}=l$ for the backward motion in the rest frame (see the tabletop experiment of Aoki et al. \cite{aoki}). In other actual experiments \cite{mugnai,budko,greaves}, however, $l_-> l_+$. The calculations for the one-way and two-way measured speeds for the corresponding gendaken experiments can be derived with the theory developed here. What is worth noticing from this class of actual experiments is that, to determine the delay time, it is assumed that the returning speed of the PE, e.g., the electric field, which is commonly guided through wires, to be a constant. In the above discussion we have seen, however, that neither the true speed nor the one-way measured speed of any PE can be determined unless we are at rest in the isotropic frame, or unless the conditions we have imposed above are satisfied. But since the earth is certainly an anisotropic frame their experimental conditions do not satisfy these requirements. This point is supported by the fact that in the anisotropic frame $\bar{u}'_-$ and $\bar{w}'_{-}$ (or the times) are function of the frame velocity $v$. To a first approximation one can consider $v$ a constant and the assumption of a constant returning speed may be acceptable for pedagogical purposes; but being strict, in accurate experiments, such assumption cannot be made. 

The ideal experimental technique to measure the one-way or true speed in the anisotropic frame is by the use of two synchronized clocks placed at the endpoints, but, an accurate synchronization process requires the knowledge of the one-way speed of light which causes a circular reasoning. This problem and other synchronization methods can be found elsewhere \cite{mansouri,feenberg,guerra1,guerra2,abreu,iyer1}.

Finally, it is important to mention that interferometric experiments like the Michelson-Morley experiment \cite{michelson1,muller,muller1,wolf} are also two-way experiments and are readily explained by the procedure we have established above; we just assume that $u=V$. Iyer and Prabhu have derived a more general expression for the one-way speed of light, they found the relation $c'=c^2(c\pm\mathbf{\hat{k}'}\cdot \mathbf{v})^{-1}$, where $ \mathbf{\hat{k}'}$ is a unit vector that points in the direction of energy flow of the light beam as determined in $K'$ and $\mathbf{v}$ is the velocity vector of $K'$ relative to $K$ and $c$ is two-way speed of light. The reader can easily verified from this equation that the harmonic mean of the speed of light, in any direction, is $c$. This fact explains the negative outcome of the experiment and, therefore, favors the existence of a unique isotropic frame.

\section{Concluding Remarks} 

One of the main goals of this contribution was simply to elucidate the physics behind the two-way measuring processes and to establish a clear distinction between the true speeds and the measured speeds. 

On the other hand, the most remarkable points that can be drawn from the two-way experimental procedure put forward here are the following. First, that the isotropic frame must exist. If this is so all other moving frames become anisotropic. Secondly, that the addition theorem of velocities derived from the TT agrees with the one-way measured values obtained from the gedanken experiment. This fact makes these transformations more suitable to describe the physics for anisotropic frames.
And thirdly, that the one-way or true speed of a PE cannot be determined in the anisotropic frame unless the experimental conditions imposed in section \ref{truespe} are fulfilled. In this respect, the present investigation was intended to boost the experimental and theoretical investigations to overcome these technical conundrums. 

\subsection*{Acknowledgments}

This article is dedicated to my colleague Dr. Nikolai Mitski\'evich from the University of Guadalajara M\'exico. I am grateful to Georgina Carrillo for helpful comments. A CONACYT grant is acknowledge.

\end{document}